\documentclass[aps,prb,twocolumn,groupedaddress,showpacs,letterpaper]{revtex4}

\usepackage[dvips]{graphics}

\begin{document}
\title{Dynamics of domain growth driven by dipolar
interactions in a perpendicularly magnetized ultrathin film}
\author{N. Abu-Libdeh} \author{D. Venus}
\email{venus@physics.mcmaster.ca} 
\affiliation{Dept. of Physics and
Astronomy, McMaster University, Hamilton Ontario, Canada}
\date{\today}

\begin{abstract}
Measurements of the ac magnetic susceptibility of perpendicularly
magnetized Fe/2 ML Ni/W(110) ultrathin films show a clear signature of
the dynamics of domain growth and domain density changes in the
striped domain pattern that this system supports.  The susceptibility
peak measured at different constant heating rates in the range 0.20
K/s $\le R \le$ 0.70 K/s shifts to higher temperature as the heating
rate is increased.  Analysis using a relaxation model demonstrates
quantitatively that the dynamics is driven by a non-equilibrium domain
density at (nearly) zero field (i.e. by dipole interactions), and that
the temperature shift is due to a response time determined by the
pinning of local domain wall segments by structural defects.  The
fundamental time scale for relaxation of the domain density driven by
dipole interactions is of order $10^{5}$ times slower than the
fundamental time scale for an individual Barkhausen step driven by an
applied field.  The increase in the fundamental time scale reflects
the relative size of dipole and Zeeman energies, and the need for the
correlated motion of the many local domain wall segments required to
affect domain growth.
\end{abstract}
\maketitle

\section{Introduction}

Ultrathin films with perpendicular anisotropy are an example of a
larger class of two dimensional systems where strong, short range
attractive interactions and weak, long range dipole repulsive
interactions lead to the formation of domain
patterns.\cite{sagui, kivelson, debell} The magnetic domain patterns in
these films provide a unique opportunity for the study of
one-dimensional domain walls (more properly domain ``lines") in a two
dimensional magnetic system and to understand the way in which domain
wall dynamics and fluctuations determine many of the finite
temperature magnetic properties of the film.

The present article outlines a quantitative experimental investigation
of the dynamics of magnetic domain growth driven exclusively by
long-range dipole interactions.  It is, perhaps, surprising that,
given the intense interest in these systems over the last few decades,
this question has not been addressed quantitatively.  The absence of
clean experiments is due to the complicated hierarchy of magnetic
relaxation processes, spanning a wide range of time scales, that
control the dynamics of these systems.  At sufficiently low
temperature, the films support an ordered stripe domain
pattern.\cite{allenspach} The motion of these existing domain walls,
when driven by an applied magnetic field, is controlled by localized
pinning of the domain walls at microstructural defects in the
film.\cite{venus} As the temperature of the films is increased, the
stripe domain density changes exponentially with temperature through
domain growth and creation mechanisms that are driven by long range
dipole interactions.\cite{portmann,won,speckmann} Finally, at a
sufficiently high temperature domain wall fluctuations can drive a
transition to a different domain pattern through the proliferation of
topological defects in the pattern itself.\cite{vaterlaus} In order to
study the domain growth dynamics driven by dipole interactions, it is
necessary to use a zero (or very low) field technique with a very wide
dynamic range.  Only then can one identify the entire heirarchy of
dynamic processes, isolate the desired mechanism and study it on the
appropriate time scale.

Almost all of the existing experimental studies do not meet these
requirements because they have used either large magnetic fields or
static imaging.  Dynamical studies have used large applied magnetic
fields either to study a magnetically saturated state very far from
equilibrium,\cite{berger} or to move domain walls using Zeeman
magnetic forces rather than weak internal dipole
forces.\cite{saratz,metaxas} Studies of the domain structure often use
magnetic microscopy in no applied field, but they are limited to a
near-static characterization of the domains.\cite{vaterlaus,won} A
very few studies infer dynamical processes qualitatively by noting the
disappearance of individual domains in successive static domain
images,\cite{portmann,portmann2} but most recognize the dynamic
limitations of magnetic microspcopy by relying on static imaging of
domain changes as a function of film
thickness,\cite{won,speckmann,vaterlaus} rather than as a function of
temperature.

In contrast, measurements of the ac magnetic susceptibility cannot
directly observe domain geometry, but do offer access a wide range of
time scales using very small magnetic fields that do not overwhelm the
dipolar interactions.  There are a handful of reports that use ac
susceptibility to characterize domain wall dynamics.  On a short time
scale, a small ac field is used to ``wiggle" existing domain walls in
order to study domain wall pinning by microstructural
defects\cite{venus,dunlavy} in the film that act on a fundamental time
of $10^{-9}$ s.  On a long time scale, measurements of the
susceptibility curve at different slow heating rates, $R \leq$ 0.1 K/s
(so that the entire curve is traced in many minutes to hours), have
revealed the dynamic processes involved in resolving topological
defects in the domain pattern itself as the system undergoes a phase
transition from one pattern to another.\cite{abulibdeh} These
collective processes act on a fundamental time of $10^0$ s.  The
present article reports experiments on Fe/2.0 ML Ni/W(110) films, and
concentrates on the intermediate time scale, where dipole interactions
drive changes in the domain density as the temperature is increased.
Measurements of the ac susceptibility curve as function of faster
heating rates, $R >$ 0.1 K/s, reveal that the peak of the
susceptibility shifts to higher temperature with greater $R$.  An
activated model of domain growth driven by the non-equilibrium domain
density at zero field reproduces these results over a range of heating
rates and sample thicknesses using a single value of a single
adjustable parameter.  This parameter, the characteristic time scale
for dipole-driven domain growth in these films, indicates a dynamics
which is $\sim 10^5$ slower than that for the field-driven motion of
existing domain walls, but much faster than that involved in removing
pattern defects.  Thus, the use of ac susceptibility as a probe of the
domain dynamics permits a coherent description of the entire hierarchy
of competing processes which determine the evolution of these systems.

The remainder of the article is divided into 4 sections. In section
II, quantitative models that describe the domain wall dynamics on
different time scales are reviewed and/or developed.  Section III
presents experimental results of ac susceptibility measurements on
ultrathin Fe/2.0 ML Ni/W(110) films, and these results are analysed
quantitively in section IV.  The final section summarizes
the findings of these studies.

\section{Theory}
\subsection{Local movement and pinning of existing domain walls: $\tau_p$}

Ultrathin (1-10 atomic layers thick) ferromagnetic films may have a
magnetocrystalline anisotropy that favours perpendicular
magnetization.  In some cases, it is strong enough to overcome
demagnetization effects, leaving a small residual perpendicular
anisotropy.\cite{bland}  The long range magnetic dipole interaction in
this geometry is antiferromagnetic and leads to the creation of
domains.\cite{yafet}  The integration of the dipole interaction over a
two dimensional film gives a logarithmic dependence of the energy on
the domain density.  Inverting this relation yields the equilibrium
domain density\cite{kashuba,abanov} $n_{eq}(T)$:
\begin{equation}
\label{neq}
n_{eq}(T) = \frac{2}{\pi \ell} \exp(-\frac{E_W(T)}{\Omega m}-1).
\end{equation}
In this equation $\ell=\pi \sqrt{\Gamma/\lambda(T)}$ is the domain
wall width, and $E_W=4 \sqrt{\Gamma\lambda(T)}$ is the domain wall
energy per unit length.  In these definitions, $\Gamma$ is the domain
wall stiffness derived from the nearest neighbour magnetic exchange
coupling, and $\lambda(T)$ is the effective perpendicular anisotropy
(including contributions from both the crystalline and demagnetization
terms).  $\Omega$ is a constant that sets the scale of the magnetic
dipole energy, and \emph{m} is the thickness of the film in
monolayers.  As the temperature is increased, thermal fluctuations
renormalize the mean field anisotropy and $\lambda(T)$ is reduced.
This causes the domain wall energy to decrease and the domain density
increases exponentially with temperature. 

The magnetic susceptibility of a domain phase is due to the motion
of existing domain walls when a small field applied perpendicular to the
surface causes the width of domains with a parallel magnetization to
grow by $\delta$ , while those with an antiparallel magnetization
shrink by $\delta$.  If the average domain density is $n$ , then the
resulting magnetization is\cite{abanov}
\begin {equation}
M = M_{sat} n\delta,
\end{equation}
where $M_{sat}$ is the saturation magnetization.  In the limit of a
small applied field, the equilibrium dc susceptibility is
\begin{equation}
\label{chieq}
\chi_{eq}(T)=\frac{4}{\pi d n_{eq}(T)} \approx A\exp (-\kappa T),
\end{equation}
where \emph{d} is the film thickness, and $A$ and $\kappa$ are
phenomenological parameters that have been shown to describe the
effect of the exponential increase in domain density in experimental
susceptibility data.\cite{venus,dunlavy}

The ac susceptibility measures the oscillation of the domain walls in a
small sinusoidal field at angular frequency $\omega$.  The response is
retarded by pinning of microscopic sections of the domain walls at
structural defects with an average binding energy $E_{pin}$.  The
response time $\tau_p$ can be modelled by an Arrhenius Law with
\begin{equation}
\label{taup}
\tau_p(T)=\tau_{0p} \exp(E_{pin}/kT).
\end{equation}
The ac magnetization is then given by a relaxation equation with
\begin{equation}
\frac{dM(t)}{dt} =\frac{-1}{\tau_p(T)} (M(t)-\chi_{eq}(T) H(t)).
\end{equation}
The steady state solution is
\begin{equation}
\label{dynamic}
\chi(T)=\frac{1-i\omega \tau_p(T)}{1 + \omega^2 \tau_p^2(T)} \: \chi_{eq}(T).
\end{equation}
The dynamic prefactor in eq.(\ref{dynamic}) describes the effect of
pinning.  The susceptibility decreases at high temperature due to the
increase in domain density, and at low temperature due to pinning.  A
peak situated roughly where $\omega \tau_p(T_{pin})=1$ divides regions
where the domain walls are pinned or free. The characteristic pinning
temperature is therefore
\begin{equation}
\label{Tpin}
T_{pin}= \frac{-E_{pin}}{k \ln(\omega \tau_{0p})}.
\end{equation}
Studies have shown that $\tau_{0p} \approx 10^{-9}$ s for this
microscopic pinning process.

\subsection{Removal of topological pattern defects: $\tau_d$}

Theoretical\cite{abanov} and numerical\cite{booth,stoycheva} studies
have indicated that a phase transition between the striped domain
phase and one of a number of delocalized domain phases is expected at
a temperature sufficently high that the domain density is large, and
domain wall fluctuations are of over-riding importance.  The
delocalized phases are characterized by the proliferation of
topological defects in the stripe domain geometry that break the
stripes into segments, and may re-orient the segments, so that
positional and/or orientational long range order are lost.

Numerical simulations\cite{bromley,cannas} have suggested that when
the system is quenched from a high temperature delocalized phase to
the low temperature striped phase, the topological defects persist for
a very long time.  This is either because macroscopic rearrangements
are required,\cite{bromley} or because the transition is
non-continuous.\cite{cannas} A recent experimental study of 1.5 ML
Fe/2ML Ni/W(110) films has observed the relaxation of the topological
defects in quenched films indirectly, through measurements of the ac
susceptibility.\cite{abulibdeh} The films were cooled from high
temperature (360 K) at a rate of $R$=-0.10 K/s, and the susceptibility
curve was measured for different constant rates of heating.  The whole
susceptibility curve was seen to shift to higher temperature when the
constant heating rate $R$ at which it was measured, was decreased.  The
long time scale (many minutes to an hour) on which this occurred,
$\tau_d(T)$, was identified as the characteristic time for topological
defects to relax.  For large $R$, the measurement time was much less
than $\tau_d(T)$, and system retained the defects of the delocalized
phase, giving a susceptibility curve with an instrinsically lower peak
temperature.  For small $R$, the measurement time was many times
$\tau_d(T)$ and the system relaxed to the equilibrium, striped phase
that had an intrinsic peak at higher temperature.  Thus the peak
temperature decreased with increasing $R$.

A quantitative description of the relaxation of topological defects
was provided by
\begin{equation}
\tau_d(T) =\tau_{0d} \exp(E_d/kT),
\end{equation}
where the fundamental time scale $\tau_{0d}$ was found to be of order
$10^0$ s, and $E_d$ is the barrier to the removal of the topological
defect.  The peak temperature as a function of the heating rate was
well described by
\begin{equation}
\label{Tpk}
T_{pk}(R)=T_0 - \Delta \exp(-t_{eff}(R)),
\end{equation}
with $t_{eff}(R)$ the effective number of time constants that have
passed during the measurement while heating at rate $R$ from initial
temperature $T_i$:
\begin{equation}
\label{teff}
t_{eff}(R)=\int_{T_i}^{T_{pk}(R)} \frac{dT}{R \tau_d(T)}.
\end{equation}
$T_0$ is the peak temperature when the relaxation to the equilibrium
stripe phase is complete. 

\subsection{Changes in domain density: $\tau_n$}

Since the \emph{entire} susceptibility curve relaxes along the
temperature axis with time constant $\tau_d$, the fundamental time
scale for changes in domain density driven by dipole interactions must
lie between the two extremes set by $\tau_{0p}$ and $\tau_{0d}$.  It
is possible to access this time scale as well, using measurments of the ac
susceptibility with small applied fields.

Measuring $\chi(T)$ involves changing the temperature at a rate $R$
(K/s).  Because structural defects pin the domain walls, it takes time
for domains to grow or contract, and the domain wall density $n(T)$
will lag behind the equilibrium value $n_{eq}(T)$ by a relaxation time
$\tau_n(T)$.  In this case, the measured susceptibility will be
\begin{equation}
\label{lag chi}
\chi(T)=\frac{1-i\omega \tau_p(T)}{1 + \omega^2 \tau_p^2(T)} \: \chi_{eff}(T),
\end{equation}
\begin{equation}
\label{chieff}
\chi_{eff}(T)=\frac{4}{\pi d n(T)},
\end{equation}
where $n(T)$ is the history-dependent present value of the domain
density during heating or cooling. It is determined by a relaxation
equation that is developed in Appendix A.  The constant rate of
temperature change, $R$, is used to remove the explict time dependence
from eq.(\ref{diffnt}), so that the relaxation equation governing
changes in domain density is
\begin{equation}
\label{diffnT}
\frac{dn(T)}{dT}=\frac{-1}{R\alpha \tau_p(T)} (n(T)-n_{eq}(T)).
\end{equation}

An important point in the development of this equation is the relation
between the time constant for the field-induced oscillation of
existing domain walls, $\tau_{p}(T)$, and the time constant for
dipole-induced changes in the domain density, $\tau_{n}(T)$.  Since
the pinning sites that retard the motion of existing domain walls when
a magnetic field is applied also impede the growth of new stripe
domains driven by dipolar interactions, the activation energies for
the two processes are the same.  However, the characteristic times
$\tau_{0n}$ is expected to be different than $\tau_{0p}$ because of
the need for correlated motion of a larger region of domain wall.  For
instance, whereas the motion of any part of an existing domain wall
will affect the response to a magnetic field equivalently, the motion
of some regions of the domain wall are much more effective than others
in changing the domain density.  Changing the domain density at
constant magnetization involves a large movement of a small length of
domain wall near the end of a domain segment.  Finally, in the
relaxation approximation, the number of active growth sites is assumed
to be proportional to the difference between the present value of the
domain density and the equilibrium value, but the value of this
proportionality constant is not obvious from first principles.  All of
these geometric factors are taken into account through an empirical
factor $\alpha$ in eq.(\ref{diffnt}), so that $\tau_{0n}=\alpha
\tau_{0p}$ and $\tau_{n}(T)=\alpha \tau_{p}(T)$.

It is straighforward to solve eq.(\ref{diffnT}) using the
phenomenological expansion for $\chi_{eq}(T)$ in eq.(\ref{chieq}).  At
high temperature the pinning is ineffective, and $n(T) \rightarrow
n_{eq}(T)$, whereas at low temperature the pinning is so effective
that $dn/dT \rightarrow 0$ and $n(T)$ saturates.  $T_n$ is a
characteristic temperature dividing these two regimes.  It can be
estimated\cite{venus2} by setting $\frac{\partial n}{\partial T} =
\frac{\partial n_{eq}}{\partial T}$ to give
\begin{equation}
n(T) = (1-R \alpha \kappa \tau_p(T))n_{eq}(T).
\end{equation}
When $R \alpha \kappa \tau_p(T_n) = 1 $ the model becomes unphysical
because the domain density cannot change quickly enough to maintain
equilibrium, and the domain density must saturate.  This occurs near the
temperature
\begin{equation}
\label{Tn}
T_n=\frac{-E_{pin}}{k \ln(R\kappa \alpha \tau_{0p})}.
\end{equation}

The ac susceptibility in eq.(\ref{lag chi}) depends on the relative
values of the characteristic temperatures for domain wall motion,
$T_{pin}$, and domain growth, $T_n$.  If $T_n < T_{pin}$, then pinning
will stop the oscillatory motion of the domain walls in the applied ac
field, even though the domain density can still change through domain
growth. The dynamical factor involving $\omega \tau_{p}$ will cut off
the susceptibility, so that it is insensitive to changes in the domain
density below $T_{pin}$.  In this case, $\chi_{eff} \approx \chi_{eq}$
over the temperature range of the susceptibility peak, and the
susceptibility will not depend upon $R$.  On the other hand, if $T_n >
T_{pin}$, then the saturation of the domain density occurs in a
temperature range where the field-induced oscillation of the domain
walls is not pinned, and the susceptibility gives a robust signal.
The curve shifts to higher temperature with increasing $R$, since
increasing $R$ increases $T_n$ in eq.(\ref{Tn}). (Note that the
direction of this shift is \emph{opposite} to that described in the
previous section for the relaxation of topological pattern defects.)
Equating $T_{pin}$ and $T_n$ gives the condition that divides these
two behaviours:
\begin{equation}
\label{alphacrit}
\alpha \kappa = \frac{\omega}{R}.
\end{equation}
Choosing measurement parameters on the right side of this equation
such that they are smaller than the combination of physical parameters
on the left side, ensures that peak temperature of the susceptibility,
$T_{pk}$, is sensitive to the relaxation of the domain density when
heating at rate $R$.  The experimentally practical range of heating
rates is from 0.20 K/s (below which the relaxation of topological
defects dominates\cite{abulibdeh}), to 1.00 K/s (above which the
decrease in measurement time at any temperature results in too much
noise).  A typical value of $\kappa$ is 0.04 $K^{-1}$.  Choosing
$\omega = 210$ Hz then gives the minimum value $\alpha_{min} \approx
10^4$ for which the measurement will be sensitive to the relaxation
dynamics of the domain density.  This magnitude for $\log_{10}(\alpha
\tau_{0p})$ is comfortably near the middle of the range between the
experimentally determined values of $\log_{10}(\tau_{0p})$ and
$\log_{10}(\tau_{0d})$.  If the actual value of $\alpha$ is smaller,
then the experimental value of $\omega$ can be reduced.

Finally, there is the question of whether or not the size of the shift
in $T_{pk}$ as a function of heating rate will be large enough to
measure.  Appendix B presents an estimate of the expected size.
According to eq.(\ref{dTdR}), the linear term in the shift of the peak
temperature of the susceptibility curve, $T_{pk}$, evaluated locally
at $R = R_0$, is given by
\begin{equation}
\label{shift}
\frac{\partial T_{pk}}{\partial R} \equiv B \approx
\frac{1}{R_0(\frac{2}{T_{pk}}+\frac{3E_{pin}}{k T_{pk}^2}+\kappa)}.
\end{equation}
From the measurements in ref.(\onlinecite{abulibdeh}) for 1.5ML Fe/2 ML
Ni/W(110) films, eq.(\ref{shift}) gives $B \approx +5$ s for $R_0=0.5$
K/s.  This small slope is consistent with the data, but cannot be
reliably detected. The parameters $E_{pin}, \kappa$ and $T_{pk}$ all
decrease quickly with film thickness.\cite{venus} Films with 2ML Fe are
estimated to have values of $B$ larger by a factor of 2 to 5.  This
should provide a clear signature of the dynamics of domain wall
growth.

\section{Experimental results}

Domain dynamics were studied in a series of Fe/2 ML Ni/W(110) films.
The nickel layer in this structure establishes a slightly strained,
epitaxial f.c.c. (111) template on the tungsten substrate, upon which
f.c.c. Fe grows epitaxially for a few layers.\cite{johnston} Due to
the f.c.c. Fe/vacuum interface, this system exhibits perpendicular
anisotropy.  Above 2.2 ML Fe, the perpendicular anistropy exists only
at low temperature and the magnetization becomes in-plane at a spin
reorientation transition.\cite{arnold} Below 1.25 ML, the magnetic
behaviour is complicated by the incomplete formation of the Fe layer.
In the present study, the film thickness was restricted to the
intermediate range where the description in section II.A has been
shown to be valid.

The films were grown in ultrahigh vacuum using thermal evaporation
from a evaporator with a calibrated internal flux monitor, following a
procedure established in previous studies.\cite{johnston} The first
monolayer of Ni was annealed to 600 K to ensure good wetting and the
growth was monitored using Auger electron spectroscopy and low energy
electron diffraction.  The substrate temperature was measured using a
W-Re5\%/W-Re10\% thermocouple embedded in the W crystal, and the
temperature was controlled by a combination of static cooling through
a copper braid attached to a liquid nitrogen reservoir, and active
heating by radiation from a filament just beneath the crystal.  The
rate of heating or cooling could be kept constant to within 0.05 K/s.
The maximum controlled cooling rate across the whole temperature range
was -0.10 K/s.  The maximum useful heating rate was 1.0 K/s, above
which the reduced number of data points introduced excessive
noise.\cite{abulibdeh2}

The ac magnetic susceptibility was measured using the polar
magneto-optic Kerr effect,\cite{arnold2} where the rotation of the plane of
polarization of linearly polarized light is proportional to the
perpendicular component of the magnetization.  A current coil near the
sample created a sinusoidal ac field of amplitude 2.0 Oe, and the very
small polarization rotation from the ultrathin film was detected using
a nearly crossed polarizer and lock-in amplification of the output
from a photodiode.  The ac field had a frequency of 210 Hz in these
studies.  The ac current through the heating filament had a much
higher frequency, so that it did not interfere with the measurements.

Because the dynamics being studied change the shape of the
susceptibility curve and shift it in temperature, considerable effort
was invested in developing procedures that removed systematic errors
and resulted in reproducible data traces.  These included: annealing
the films to 400K before measurements commenced, and subsequently
never heating above 360K; cooling the sample from 360 K at a rate of
-0.10 K/s before each susceptibility measurement was recorded during
heating; discarding the first heating trace to ensure a consistent
magnetic history for each measurement; randomizing the order of the
measurements for different values of heating rate; demonstrating that
the lock-in time constant of 2 s had no effect on the shape or
position of the susceptibility curve for the range of heating rates
used.

\begin{figure} 
\scalebox{.4}{\includegraphics{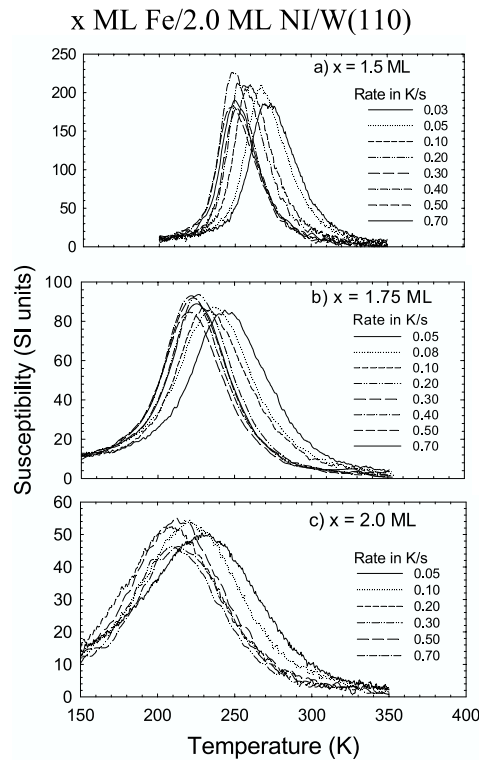}} 
\caption{Magnetic ac susceptibility of x ML Fe/2 ML Ni/W(110) films as
a function of temperature for different constant heating rates.  All
the measurements were performed after cooling the sample from 360 K at
a rate of -0.10 K/s.  An ac field of 2.0 Oe at a frequency of 210 Hz
was applied normal to the film.  Parts a), b) and c) show sequential
measurements from a single film of thicknesses 1.5 ML, 1.75 ML and 2.0
ML of Fe, respectively. }
\end{figure}

Figure 1 shows the real part of the susceptibility curves measured for
three separate films with Fe thicknesses of 1.5 ML, 1.75 ML and 2.0 ML
(all $\pm$0.1 ML), and heating rates varying from 0.03 K/s to 0.70
K/s.  (Note: the data for 1.5 ML Fe is the same data that is analysed
in ref.\onlinecite{abulibdeh}).  The absolute scale of the susceptibility
is uncertain within a factor of about 2 because the magneto-optic
Voigt parameter is not well known for these films.  However, the
relative scale for all traces on all plots is consistently calibrated
to the absolute optical rotation.  The general shape of the curves is
consistent with many previous studies.  At high temperature the
susceptibility decreases exponentially as the domain density
increases.  At low temperature the susceptibility decreases
exponentially as the domain walls become pinned in structural
defects. The peak in the susceptibility occurs at an intermediate
temperature due to the interrelation of these two processes, and
depends strongly on the film thickness.

\begin{figure} 
\scalebox{.4}{\includegraphics{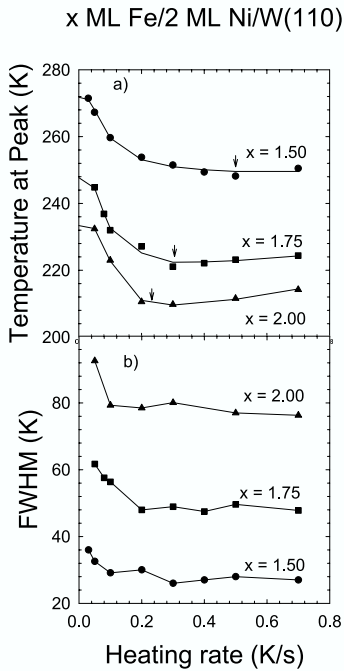}} 
\caption{a) The temperature at the peak of the susceptibility curve,
$T_{pk}$, as a function of the heating rate, for films of different Fe
thickness.  The arrows mark approximately the heating rate at which
$T_{pk}$ has a minimum.  The fitted lines are discussed in section IV.
b) The full-width at half-maximum (FWHM) of the susceptibility curves as a
function of heating rate.  Lines simply connect the points.}
\end{figure}

The dependence of the susceptibility on the heating rate that is
evident in fig. 1 is summarized more clearly in fig. 2 using two
quantities: the temperature at the peak of the susceptibility,
$T_{pk}$, and the full width at half maximum of the peak.  The
systematic variation of these quantities with heating rate is the
subject of section IV.  The remainder of this section is concerned
with comparing the variation of the susceptibility with film thickness
to previous findings. Quantitative fits to the data, using
eq.(\ref{dynamic}) with parameters defined in eq.(\ref{taup}) and
(\ref{chieq}), are shown in fig. 3.  Part a) demonstrates the quality
of the fits and confirms that the essential points of the model are
valid.  Values of the parameters $E_{pin}$ and $\kappa$ are given in
parts b) and c), respectively.  As has been shown
previously,\cite{abulibdeh} the average pinning energy, which is a
structural property of each film, is independent of the heating rate.
The pinning energy decreases rapidly with film thickness.  A previous
study\cite{venus} has shown that the pinning is due to changes in the
perpendicular anisotropy caused by changes in the thickness at
monolayer steps in the film, and that $E_{pin} \sim d^{-3/2}$ for
measurements of a single film grown sequentially to a number of
thicknesses.  The current results are qualitatively consistent with
this finding, but quantitative comparisons are not possible among a
series of independently grown films with unrelated microstructure.
The parameter $\kappa$ depends on both film thickness and the heating
rate.  The variation with thickness that displaces the curves one from
another in fig. 3c is in qualitative agreement with eq.(\ref{chieq}).
Because the surface anisotropy varies as $1/d$, $\kappa$ can be
ultimately derived from eq.(\ref{neq}) as $\kappa \approx \frac{1}{d}
\frac{\partial E_W}{\partial T} \sim d^{-3/2}$.

\begin{figure} 
\scalebox{.4}{\includegraphics{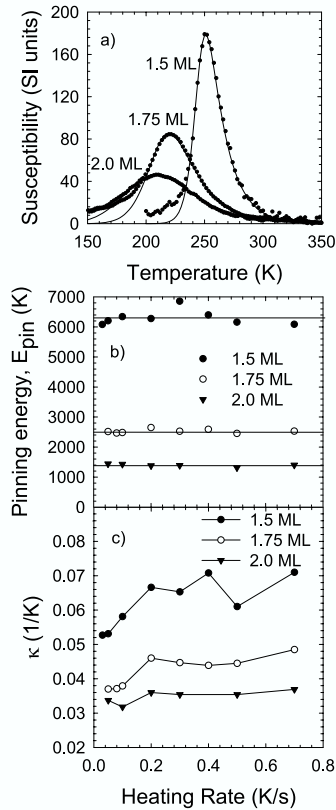}} 
\caption{Fits to the data using the phenomenological parameters from
eq.(\ref{dynamic}). a) Representative fits for measurements at a
heating rate of 0.30 K/s for different film thicknesses.  b) The
pinning energy for local Barkhausen steps of microscopic segments of a
domain wall.  c) The paramter $\kappa$, which reflects the exponential
increase in domain density with temperature.  Lines simply connect the
points.}
\end{figure}

The thickness dependence of $E_{pin}$ and $\kappa$ leads to the
systematic displacement of the susceptibility curves with thickness
observed in fig. 2.  A gross measure of $T_{pk}$ is the pinning
temperature $T_{pin}$, and eq.(\ref{Tpin}) indicates that this scales
with $E_{pin}$.  Thus the curves in fig. 2a are displaced to lower
temperature for thicker films.  Similarly, since the two exponential
factors ($E_{pin}$ and $\kappa$) that cut off the susceptibility above
and below its peak value vary inversely with a power of the thickness,
the peak width in fig. 2b increases as the film thickness increases.

\section{Analysis and Discussion}

As the heating rate increases, $T_{pk}$ first decreases sharply, and
then reverses and gradually increases.  The value of $R$ at which this
reversal occurs (marked by the arrows in fig. 2a depends
systematically on the film thickness.  This suggests that there are
two competing dynamical processes -- one that dominates at small $R$
and the other that dominates at large $R$.  It is important to note
that the changes in the peak amplitude with $R$ in fig. 1 are very
modest even as the peak temperature changes substantially.

The rapid decrease in $T_{pk}$ and in the peak width at small $R$ are
correlated to an increase in $\kappa$ in fig. 3. This behaviour was
analysed in ref. \onlinecite{abulibdeh}, where it was related to the slow
relaxation of topological defects in the domain pattern after the
sample is quenched from a high temperature, delocalized phase, as in
section IIB. The present data now confirms this observation at other
thicknesses.  The second, less dramatic, process at large $R$ is more
obvious in the new data at thicknesses of 1.75 and 2.00 ML.  According
to the analysis in section IIC, the increase in $T_{pk}$ at large
values of $R$ is qualitatively consistent with the dynamics of domain
density changes and growth on an intermediate time scale.  This
agreement suggests a quantitative analysis using a combination of
eq.(\ref{Tpk}) and (\ref{shift}):
\begin{equation}
\label{combined}
T_{pk}(R)=T_0 - \Delta \exp(-t_{eff}(R)) + BR. 
\end{equation}

\begin{figure} 
\scalebox{.4}{\includegraphics{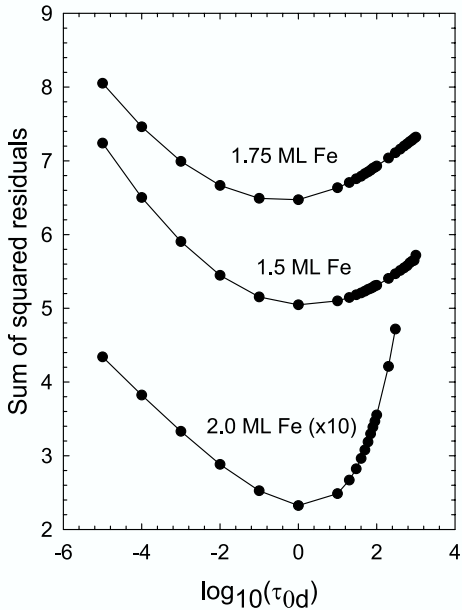}} 
\caption{The sum of squared residuals for the independent fits of
eq.(\ref{combined}) to the peak temperatures as a function of heating
rate in fig. 2a.  The residuals for the film with 2.0 ML Fe are
multiplied by 10 to place them on the same plot.  The residuals for
this film are smaller because there are data points at fewer heating
rates.}
\end{figure}

Figure 4 presents the least-squares residuals of the fit of the
experimental peak temperatures to this expression by varying $T_0,
\Delta, E_d$ and $B$ for a range of values of the time constant
$\tau_{0d}$.  A consistent optimum value of $\log_{10} \tau_{0d} =
0.00\pm 0.05$ is found.  Table I gives the best fit parameters for
each film thickness.  Error estimates on these parameters, as well as
on $\tau_{0d}$, are derived by holding all the other parameters
constant, and finding the range of variation that changes the squared
residuals by 1.0. The fitted curves are superimposed upon the data in
fig. 2a.

\begin{table}
\begin{center}

\begin{tabular}{|c|c|c|c|c||c|} \hline

Fe ML   &  $E_d$ (K)  &  $T_0$  (K) &  $\Delta$ (K) 
&  $B$ fit (s)  & $B$ calc. (s) \\ \hline 
1.50 & 1560$\pm$25 & 271.8$\pm$0.6 & 26.0$\pm$0.9 
& 2.7$\pm$1.8 & 3.95$\pm$0.5 \\ \hline
1.75 & 1390$\pm$13 & 247.7$\pm$0.5 & 33.8$\pm$0.8
& 11.0$\pm$1.5 & 8.91$\pm$0.5\\ \hline
2.00 & 1270$\pm$12 & 234.1$\pm$).5 & 38.1$\pm$0.7
& 21.0$\pm$1.5 & 16.1$\pm$0.5 \\ \hline

\end{tabular}
\caption{Parameters for fitting the data in fig. 2a to
eq.(\ref{combined}) appear to the left of the double line.  In all
cases $\log_{10}(\tau_{0d}) = 0.00\pm0.05$. The parameter $B$ calc. to
the right of the double line is determined in fig. 5.}
\end{center}
\end{table}

These results confirm the earlier analysis of the data for the 1.5 ML
Fe films,\cite{abulibdeh} where the term $BR$ due to to the dynamics
of domain density changes was neglected.  Since $B$ is found to be
small for this thickness, the values of the other parameters are
essentially unchanged from those found in the previous study.  The new
data for thicker films show that, upon quenching from high
temperature, the relaxation parameter $\tau_{0d}$ and $E_d$ for
topological defects in the domain pattern do not depend strongly on
the film thickness.  It continues to take many minutes to an hour to
remove these defects and for the magnetic susceptibility to take the
equilibrium curve representing the ordered stripe phase.  The fact
that $E_d$ does not scale with $E_{pin}$ further substantiates the
conclusion that the removal of the topological pattern defects is not
limited by microscopic pinning mechanisms, but rather by the low
probability that weak dipole interactions will drive the co-ordinated
domain wall fluctuations required to make the mesoscopic changes
required to remove the defect from the pattern. For all thicknesses,
the defect relaxation is reflected as well in a decrease in $\kappa$.
As was previously noted,\cite{abulibdeh} the larger value of $\kappa$
in the presence of topological defects may represent their influence
on the non-equilibrium free energy, creating an increase in the
magnetic ``stiffness" as measured by the susceptibility.

Because the constant $B$ is not small for the 1.75 and 2.00 ML Fe
films, the data cannot be reasonably fit without this term.  This
proves that even as the topological defects are relaxing, a second,
faster relaxation mechanism is present in these films.  The sign and
order of magnitude of the second effect is consistent with the
estimates from eq.(\ref{shift}) for relaxation of the domain
density, and indicate that the domain density and topological defect
density are not strongly coupled, but relax rather independently.  

\begin{figure} 
\scalebox{.4}{\includegraphics{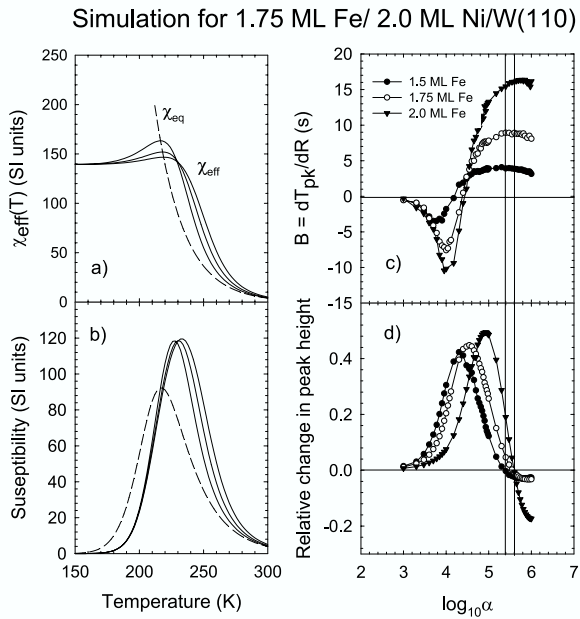}} 
\caption{Calculation of the changes in the susceptibility peak due to
domain growth dynamics for a film with 1.75 ML Fe.  a) Parameters fit
in fig. 2 for the film with a heating rate of 0.30 K/s are used to
calculate $\chi_{eq}$ directly (dashed line), and $\chi_{eff}$ through
the integration of eq.(\ref{diffnT}) (solid lines).  The heating rates
are (left to right) 0.20, 0.40 and 0.70 K/s.  The single adjustable
parameter $\alpha = 10^{5.5}$.  b) The results in part a) are
multiplied by the dynamic prefactor in eq.(\ref{dynamic}) to simulate
the measured susceptibility.  c) Calculations of the susceptibility as
a function of $R$, such as those in part b), are used to find the
linear dependence of $T_{pk}$ on $R$ as as function of $\alpha$.  d)
Calculations of the susceptibility as a function of $R$, such as those
in part b), are used to find the relative change in the peak amplitude
as a function of $\alpha$. }
\end{figure}

The interpretation of the constant $B$ can be tested quantitatively.
Fig. 5 gives an example calculation for the film with 1.75 ML Fe. The
parameters $A, \kappa, \tau_{0p}$ and $E_{pin}$ are those determined in
fig. 3a by the fit to eq.(\ref{dynamic}) for the data with $R$=0.30.
In fig. 5a the result for $\chi_{eq}(T)$ derived from $A$ and $\kappa$
is given by the dashed line.  The same parameters are then used to
evaluate $n_{eq}(T)$ and derive $n(T)$ as solutions to
eq.(\ref{diffnT}) for different heating rates.  The initial condition
for each is the saturated, ``frozen'' domain density that results from
solving eq.(\ref{diffnT}) as a function of cooling from the
equilibrium state at 350 K, at a rate of $R$=-0.10 K/s.  Once $n(T)$
is known, $\chi_{eff}(T)$ is easily derived from
eq.(\ref{chieff}). The solid curves in fig. 5a give $\chi_{eff}(T)$
for heating rates of 0.20, 0.40, and 0.70 K/s (moving from left to
right in the figure), for the choice $\alpha = 10^{5.5}$.  It is clear
that $\chi_{eff}(T)$ peaks at lower temperature when $R$ is smaller,
since the system has more time to relax toward the equilibrium state.
Fig. 5b presents $\chi(T)$ by multiplying the results of part a) by
the dynamical pre-factor defined in eq.(\ref{dynamic}) and (\ref{lag
 chi}).  It can be seen that the calculated susceptibility shifts
gradually to higher temperature with higher $R$, but that the
amplitude of the peak changes very little.

The position $T_{pk}$ and peak amplitude can be taken from a series
of curves such as those in fig. 5b.  The linear dependence of these
quantities as a function of $R$ can be extracted to give one point on
each of fig. 5c and 5d.  Repeating the calculation for many values of
$\alpha$ produces the entire plots.  The three traces on these plots
correspond to parameters fit to the data in fig. 3a for thicknesses of
1.5, 1.75 and 2.0 ML Fe.  As was previously discussed, for small
values of $\alpha$, the characteristic temperature for the freezing of
the domain density, $T_n$, is much less than that for the pinning of
the domain wall segments, $T_{pin}$, and since any changes in domain
density dynamics with $R$ cannot be observed, $B \approx 0$.  For
$\alpha$ < $10^5$, $T_{pin} << T_n$, and $T_{pk}$ depends upon $R$
through a relatively constant, non-zero linear coefficient $B$.  In
the intermediate range of $\alpha$, $T_{pin} \approx T_n$ and the two
dynamic mechanisms interact in a complex manner that gives a strong
sensitivity of $B$ to the value of $\alpha$.

The two vertical lines on fig. 5c and 5d highlight the region near
$\alpha = 10^{5.5}$.  In this region, $B$ is positive and relatively
insensitive to $\alpha$, and the peak height depends only weakly on
the heating rate, in agreement with the experimental data.  These
constraints determine the value of $\alpha$ for these films. The
corresponding calculated values of $B$ taken from fig. 5c are appended
to Table I, under the heading ``B calc.''.  It can be seen that this
single value of the single adjustable parameter $\alpha$ gives good
quantitative agreement with ``B fit'' (within limits of uncertainty)
determined from the independent analysis of the plots of $T_{pk}$ in
fig. 2.

These results are strong support for the identification of the
dynamics of the domain density as the cause of the gradual rise in
$T_{pk}$ at large heating rates.  These findings validate the model
presented in appendix A, where the domain density changes by domain
growth driven by dipole interactions, but ultimately hindered by the
same actviated domain wall pinning as the local movement of domain
walls in response to an applied field. The value of $\alpha \sim
10^{5.5}$ derived from the analysis demonstrates that the domain
density relaxes at an intermediate time scale, that is $10^{5.5}$
times slower than the time scale for local pinning of a segment of a
domain wall, but much faster than that required for the relaxation of
topological defects in the domain pattern. 

\section{Conclusions}

Magnetic susceptibility measurements of Fe/2 ML Ni/W(110) films show
the effects of three separate magnetic relaxation mechanisms that span
a wide range of fundamental time scales.  Two of these are readily
apparent and have been studied previously.  Field-driven oscillatory
movement of existing domain walls has a fundamental time scale of
$10^{-9}$ s and is pinned by structural defects in the film.  This
dynamical response cuts off the susceptibility below a pinning
temperature $T_{pin}$ and gives the ac susceptibility its characteristic
shape.  Topological domain pattern defects that persist after
quenching from a delocalized phase are removed on a fundamental time
scale of $10^0$ s, and cause large (25 to 40 K) shifts in the entire
susceptibility curve when it is measured at a slow heating rate ($R
\le 0.10$ K/s).  This shift is to lower temperature as $R$ is
increased.  Through a systematic study of these effects, it has been
possible in the present article to isolate and study a third relaxation
mechanism at an intermediate time scale.  This is the dipole-driven
changes in domain density that create a smaller shift of the ac
susceptibility curve to higher temperature as $R$ is increased ($R \ge
0.20$ K/s).  The increase in characteristic times of the mechanisms
reflects the increase of the size of feature upon which each mechanism
works -- from the magnetic response of individual Barkhausen steps, to
the correlated fluctutations required to drive the growth of domains,
to the even more complex rearrangements needed to resolve topological
defects involving multiple domains.

A simple activated relaxation model of domain density dynamics
provides a quantitative description of 22 susceptibility traces
measured at different heating rates on films with three different
thicknesses, using a single value of a single adjustable parameter.
This parameter is $\alpha$, the proportionality between the relaxation
time for local domain wall pinning and domain density changes,
$\tau_n(T) = \alpha \tau_p(T)$.  The functional form of this relation
gives strong support for three features of the model.  Most
fundamentally, it confirms that the changes in domain density in the
experiment are driven by the departure of $n(T)$ from $n_{eq}(T)$ in
the absence of a magnetic field, that is by long range dipole forces.
Second, that the same structural defects that pin local domain wall
movement limit the growth of domains.  Finally, that the nucleation of
domains is not a limiting factor in changing the domain density in
these experiments, and can be neglected.  This could be because the
nucleation energy is less than the pinning energy, or due to the
magnetic history of the films.  Since all films are grown at high
temperature, they start out with a large density of domains.  Upon
cooling, the domain density decreases by shrinking the domains, but
may leave behind small nuclei which are not eliminated by the weak
dipole forces.\cite{saratz} Upon heating, growth can occur at the
existing nuclei.

The quantitative value $\alpha \approx 10^{5.5}$ confirms that the
domain density relaxes on an intermediate time scale.  The separation
of the three relaxation mechanisms by orders of magnitude permits them
to be studied independently by appropriately designed experiments.  In
particular, changes in the domain density are found to occur
independently from the resolution of topological defects, both because
of the very different time scales, and the unrelated activation
energies.  This statement does not mean that structural defects do not
affect the local evolution of domain walls in the topological defects,
but rather that they are not the limiting factor in their dynamics.
These findings are in essential agreement with microscopy studies.
The very long relaxation times for topological defects permits them to
be imaged, and spatial Fourier transforms of images show a loss of
orientational order.  The much quicker changes in stripe density,
however, cannot be followed, except through the sudden disapperance of
individual stripe domains between image
line-scans.\cite{portmann,portmann2}

The present experiments have many similarities to simulations by
Bromley \emph{et al.}.\cite{bromley} These authors also identify three
dynamical regimes with different time scales for the relaxation of the
magnetic state of Ising spins on a square lattice.  There are also
important differences, since the simulations study relaxation from a
magnetically saturated state at constant temperature, rather than
quenching and heating at (almost) zero field.  Nonetheless, the
initial, fast relaxation of the magnetization in the simulations,
driven by Zeeman energies once the field is removed, is clearly
analogous to the field-driven domain wall oscillation in the current
experiments.  Similarly, the subsequent growth and connection of
domain segments to form stripes on an intermediate time scale is very
reminescent of the model used in the present article for the change
the domain density.  In the simulations, this process is driven both
by dipole interactions and a residual non-zero magnetization.
Finally, Bromley \emph{et al.}\cite{bromley} find that topological
defects form boundaries separating regions of different stripe
orientation, and require a very long time to relax.  This is
qualtitatively similar to the relaxation of topological defects
observed indirectly here.  A comparative analysis should not be
pressed too far, since the simulations are for very narrow Ising
domain walls and do not include a mechanism for domain wall pinning.

\section{Acknowledgements}
We are thankful for the continuing technical assistance of M. Kiela,
and for insightful discussions with J.P. Whitehead and A.B.  MacIsaac.
This work was supported by the Natural Sciences and Engineering
Research Council of Canada.

\appendix
\section{}
In this appendix, a relaxation equation for changes in the domain
density is developed.  Assume that the long domains characteristic of
the stripe phase are aligned along the y-axis, and that the
measurement samples an area of dimension $L_x\times L_y$.  If the
$i^{th}$ domain segment has length $\mathcal{L}_i$, then the average
domain density is the total domain perimeter divided by the sample
area:

\begin{equation} 
n=\frac{1}{L_xL_y}  \sum_i 2\mathcal{L}_i . 
\end{equation} 

The rate of change of the domain density is then 
\begin{equation} 
\frac{dn}{dt}=\frac{2}{L_xL_y} \sum_i \frac{d\mathcal{L}_i}{dt} = 
\frac{2}{L_xL_y} \sum_i 2v_i. 
\end{equation} 
$v_i$ is the velocity of the growing (or contracting) tip of the domain 
segment.  The factor of 2 enters because the length of a domain segment 
changes at both ends.  

Due to pinning, the activated average velocity is\cite{ferre}

\begin{equation} 
\label{velocity} 
v=v_0 \exp(\frac{-(E_{pin}-E_{dipole})}{kT}). 
\end{equation} 
The fundamental speed $|v_0|$ is given by the product of the
Barkhausen step size $s$ and the ``attempt'' frequency $\nu_0 =
1/\tau_{0p}$. The structural defects that pin the domain walls and
retard the growth of the domain segments are the same ones that pin
the domain walls and retard the change in the width of the domains
when an ac field is applied.  Therefore, $E_{pin}$ and $\tau_{0p}$ are
the same quantities in eq.(\ref{taup}) and (\ref{velocity}).  Given
that these pinning energies are of order $10^3$ K, and that the dipole
energy $E_{dipole}$ driving domain formation is of order $10^0$ K, the
latter can be neglected in eq.(\ref{velocity}).  To the extent that
the distribution of pinning energies can be represented by an average
energy $E_{pin}$, then the distribution of domain wall velocities
$v_i$ can be represented by the average $v$.

The sum $\frac{1}{L_xL_y} \sum_i 2$ represents the density of active
growth fronts. In a relaxation approximation, the density of active
growth fronts is proportional to the total density of growth fronts,
$\rho$, times the deviation of the number of domain walls from its
equilibrium value, $(N-N_{eq})$. The total density of growth fronts
includes the ends of existing domain segments as well as latent
domains nucleated at defects that may grow if the conditions are
favourable.  This model does not include an explicit nucleation
energy, but rather assumes that, upon cooling, the weak dipole energy
differences can reduce the domains centred on the nucleation centres
to a size such that the domain perimeter is negligible compared to
that of the remaining domain segments, but cannot completely remove
them. This assumption is supported by microsopy studies of domain
formation.\cite{saratz}

Putting these factors together,
\begin{equation}
\frac{dn}{dt}=-\frac{2\beta s\rho L_x}{\tau_{0p}}\exp(\frac{-E_{pin}}{kT})
(n-n_{eq}).
\end{equation}
The proportionality constant, $\beta$, between the number of active
growth sites and $\rho (N-N_{eq})$, is difficult to derive.  It is
related to how strongly the driving dipole energy depends upon the
number of domain walls.  To the extent that dipole energies are
``small'' and depend only logarithmically on the number of domains,
$\beta$ is expected to be ``small'' and produce a long relaxation
time.  The factor $1/(2 \rho L_x)= \xi$, the average separation of
growth sites in the y direction.  Finally, the differential equation
describing the relaxation of the domain density is
\begin{equation}
\label{diffnt}
\frac{dn(T)}{dt}=\frac{-1}{\tau_n(T)} (n(T)-n_{eq}(T)),
\end{equation}
where $\tau_n=\alpha \tau_p$, with $\alpha = \xi/(s \beta)$.

\section{}
To find how the temperature of the peak of the susceptibility,
$T_{pk}$, depends upon the heating rate, it is first necessary to find
an expression for $T_{pk}$.  Starting from eq.(\ref{lag chi}), the
derivative $\frac{\partial \chi}{\partial T}$ is set to zero to yield
the implicit relation
\begin{equation}
\label{deriv}
\frac{2E_{pin}}{kT_{pk}^2} \frac{\omega^2
\tau_{p}^2(T_{pk})}{1+\omega^2 \tau_{p}^2(T_{pk})} =
\frac{1}{n(T_{pk})} \left. \frac{\partial n(T)}{\partial T} \right
|_{T=T_{pk}}.
\end{equation}
Equation \ref{diffnT} is substituted in for $\frac{\partial
n}{\partial T}$, recalling the relation $\tau_n(T)=\alpha
\tau_{p}(T)$.  This gives
\begin{equation}
\label{implicit}
\frac{2E_{pin}}{kT_{pk}^2} \frac{R \alpha}{\omega} \frac{\omega^3
\tau_{p}^3(T_{pk})}{1+\omega^2 \tau_{p}^2(T_{pk})} =
\frac{n_{eq}(T_{pk})}{n(T_{pk})}-1.
\end{equation}
As was discussed in section IIC, $T_{pk}$ will depend on the
relaxation of the domain density only when $T_n > T_{pin}$.  In this
case $\omega^2 \tau_{pin}^2(T_{pk}) << 1$ and can be neglected in the
denominator of the dynamical factor.  A further consequence is that
near $T_{pk}$, $n$ is substantially less than $n_{eq}$, as can be seen
in fig. 5a and 5b.  Using these approximations gives as result that is
not valid in the limit of very small $R$:
\begin{equation}
\label{approx}
\frac{2E_{pin}}{kT_{pk}^2} \frac{R \alpha}{\omega} \omega^3
\tau_{p}^3(T_{pk}) \approx \frac{n_{eq}(T_{pk})}{n(T_{pk})}.
\end{equation}

To find the implicit derivative $\frac{\partial T_{pk}}{\partial R}$
of the left hand side of this equation is straight forward.  For the
right hand side,
\begin{eqnarray}
\lefteqn{\frac{\partial}{\partial
R}\left(\frac{n_{eq}(T_{pk})}{n(T_{pk})}\right) =
\left(\frac{n_{eq}(T_{pk})}{n(T_{pk})}\right) \times} \nonumber \\
& & \times \left(\kappa -\frac{1}{n(T_{pk})}
\left. \frac{\partial n(T)}{\partial T}\right|_{T=T_{pk}}\right)
(\frac{\partial T_{pk}}{\partial R}),
\end{eqnarray}
where the ansatz in eq.(\ref{chieq}) has been used.

Substituting in this equation from eq.(\ref{deriv}) and
(\ref{approx}), and setting the derivatives of the left and right hand
side of eq.(\ref{approx}) equal, yields after rearrangement
\begin{equation}
1 \approx (\frac{\partial T_{pk}}{\partial
R})\left(\frac{2}{T_{pk}}+\frac{3E_{pin}}{kT_{pk}^2}(1-\frac{2}{3}\omega^2
\tau_p^2(T_{pk}))+\kappa\right)R.
\end{equation}
Using once again the condition $\omega^2 \tau_{pin}^2(T_{pk}) << 1$
gives the final estimate
\begin{equation}
\label{dTdR}
\frac{\partial T_{pk}}{\partial R} \approx
\frac{1}{R(\frac{2}{T_{pk}}+\frac{3E_{pin}}{k T_{pk}^2}+\kappa)}.
\end{equation}

\end{document}